\documentclass[a4paper, 12pt]{article}
\usepackage{jheppub}
\pdfoutput=1

\usepackage{graphicx, subfigure}  % needed for figures
\usepackage{dcolumn}   % needed for some tables
\usepackage{bm}        % for math
\usepackage{amssymb}   % for math
\usepackage{amsmath} 
\usepackage{mathalfa}
\usepackage{braket} % for kets
\usepackage[british]{babel}
\usepackage[applemac]{inputenc}
\usepackage{bbold}

\usepackage{empheq}

\newcommand{\be}{\begin{equation}}
\newcommand{\ee}{\end{equation}}
\newcommand{\bea}{\begin{eqnarray}}
\newcommand{\eea}{\end{eqnarray}}

\newcommand{\Pv}{\ensuremath{\vec{\Pi}}}

\title{Universal Properties of Type IIB and F-theory  Flux Compactifications at Large Complex Structure}

\author[1]{M.C.~David Marsh,}
\author[2,3]{Kepa Sousa}
\affiliation[1]{Department of Applied Mathematics and Theoretical Physics,
University of Cambridge, Cambridge, CB3 0WA, United Kingdom}
\affiliation[2]{Instituto de Fisica Teorica UAM-CSIC
Universidad Autonoma de Madrid, Cantoblanco, 28049 Madrid, Spain}
\affiliation[3]{Department of Theoretical Physics and History of Science,
University of the Basque Country UPV/EHU, 48080 Bilbao, Spain}
\emailAdd{m.c.d.marsh@damtp.cam.ac.uk}
\emailAdd{kepa.sousa@csic.es}

\abstract{ 
 We consider flux compactifications of type IIB string theory and F-theory in which the respective superpotentials at large complex structure are dominated by cubic or quartic terms in the complex structure moduli. In this limit, the low-energy effective theory exhibits universal
 properties that are insensitive to the details of the compactification manifold or the flux configuration. 
 Focussing on the complex structure and axio-dilaton sector, we show that there are no vacua in this region and the spectrum of the Hessian matrix is highly peaked and consists only of three distinct eigenvalues ($0$, $2m_{3/2}^2$ and $8m_{3/2}^2$), independently of the number of moduli. We briefly comment on how the inclusion of K\"ahler moduli 
  affect these findings. Our results generalise those of  Brodie \& Marsh \cite{Brodie:2015kza}, in which these universal properties   were found in a subspace of the large complex structure limit of type IIB compactifications. 

}

\date{\today}

\begin{document}

\maketitle
\section{Introduction}

String theory can be compactified on any of a large set of permissible manifolds,  each typically admitting numerous configurations of fluxes and branes. At low energies, this results in a large set of string theory-derived  four-dimensional effective field theories.  
Unfortunately, typical examples of such theories involve a large number of fields and are very challenging to explicitly construct and solve.  Statistical methods that circumvent direct enumeration of explicit solutions have proven valuable \cite{Ashok:2003gk},  
however, as such methods invariably simplify the problem by making certain approximations, 
it is of obvious importance to delineate the corresponding regimes of applicability.

The primary assumption of statistical studies of flux compactifications is that 
the generalised Dirac quantisation condition on the fluxes can be ignored when the typical number of flux quanta is large.
In other words, the statistical properties of the ensemble of effective theories obtained from summing over quantised fluxes
are approximated by a continuous integral over these fluxes so that, schematically, 
\be
\sum_{\tilde N
} \longrightarrow \int {\rm d} \tilde N \, , \label{eq:contFlux}
\ee
for the \emph{a priori} quantised flux $\tilde N$.
This \emph{`continuous flux approximation'} is of key importance in the derivation of the index  of supersymmetric flux vacua \cite{Ashok:2003gk}, as well as the subsequent statistical studies \cite{Denef:2004dm, Denef:2004ze, Denef:2004cf}.

Building on the continuous flux approximation, it was pointed out in \cite{Denef:2004cf} that the distributions of certain matrices that appear in the low-energy effective theory, should -- for sufficiently complicated compactification manifolds with many moduli -- be well approximated by models from random matrix theory (RMT).  
This idea was further expounded upon in \cite{Marsh:2011aa}, where it was shown that in such `random supergravities', typical de Sitter critical points are exceedingly unlikely to be metastable vacua. The celebrated statistical universality of RMT models suggests that these results should be roughly independent of the microscopic details of the distributions of the couplings in the theory. 

However, when the distributions of the couplings are substantially modified, the appropriate  %RMT 
model to describe the ensemble of effective theories may be different, and the nature of these  `universal' limits could change.   For instance,   supergravity theories with multiple sectors   appear frequently in string compactifications and may support de Sitter vacua at a much higher frequency  than typical critical points in random supergravity   \cite{Marsh:2011aa, Sousa:2014qza,Marsh:2014nla, Achucarro:2015kja}. Further discussions of random supergravities and string compactifications can be found in    \cite{Chen:2011ac, Danielsson:2012by, Bachlechner:2012at,   MartinezPedrera:2012rs,Bachlechner:2014rqa,Long:2014fba}.

Recently, it was shown that  the low-energy theory of type IIB flux compactifications close to a large complex structure (LCS) point satisfies certain universal properties that are independent of the number of moduli, the details of compactification geometry, or the detailed flux choice \cite{Brodie:2015kza}. Specifically, upon restricting to the complex structure and axio-dilaton sector of compactifications with   $h^{1,2}$ complex structure moduli, it was shown that an $h^{1,2}+3$ real-dimensional subspace of the LCS limit contains no supersymmetric vacua and that the spectrum of the Hessian at any point in this region  
is given by,
\begin{empheq}[box=\fbox]{align}
\vspace*{.5 cm}
m^2_{i} =
\left\{
 \begin{array}{l c l }
 0 &~~& i=1, \ldots,h^{1,2}, \\
  2\, m_{3/2}^2 &~~& i=h^{1,2}+1, \ldots, 2h^{1,2}+1, \\
  8\, m_{3/2}^2 && i= 2(h^{1,2}+1) \, , 
\end{array}
\right.
\label{eq:spect}
\vspace*{.5 cm}
\end{empheq} 
where $m_{3/2} = e^{K/2} |W|$ denotes the gravitino mass.  These results are in stark contrast to the expectations derived from the continuous flux approximation and the proposed RMT models (which predict a smooth spectrum), and thus serve to indicate a limit of applicability of  these statistical techniques and approximations. Moreover, it was shown in \cite{Brodie:2015kza} that the scalar potential 
in this region of the moduli space is too steep to support inflation, thus presenting a significant obstacle for realising inflation at large complex structure.
 
In this paper, we extend the results of \cite{Brodie:2015kza} to the entire $2(h^{1,2} +1)$ dimensional 
moduli space  for flux compactifications of type IIB string theory at large complex structure. We show that  as long as the superpotential is dominated by terms cubic in the complex structure moduli,  the spectrum of  
the Hessian is given by equation \eqref{eq:spect}.

The generality of this conclusion is intimately related to the unbounded relative growth of a single period of the compactification manifold close to a LCS point. Our proof relies heavily on the universal form of the orthonormal-frame Yukawa couplings in the LCS limit \cite{Denef:2004ze,Dimofte:2008jg,Farquet:2012cs}, from which all the relevant properties of the low-energy theory can be derived.

Flux compactifications of F-theory are particularly interesting since  the topological complexity of elliptically fibered Calabi-Yau four-folds greatly exceeds that of three-folds, and one may expect that this should lead to a very large number of flux vacua  arising from F-theory compactifications  \cite{Taylor:2015xtz}. Here, we show that our results for type IIB flux compactifications generalise in a straightforward way to a corresponding LCS limit of F-theory compactifications. In particular, we show that in regions of the moduli space where the superpotential is dominated by terms quartic in the complex structure moduli, there are no vacua, and the spectrum of complex structure deformations  is again given by equation \eqref{eq:spect}, \emph{mutatis mutandis}.

 Finally,  we would like to emphasise that our derivation neglects possible  interactions of the axio-dilaton and complex structure moduli with other degrees of freedom, such as the K\"ahler moduli.  Including  such  interactions will generically affect the spectra, unless the additional sectors can be integrated out or consistently truncated while preserving  supersymmetry \cite{Gallego:2008qi,Gallego:2009px,Brizi:2009nn,Binetruy:2004hh,Achucarro:2007qa,Achucarro:2008sy,Achucarro:2008fk}.   In string compactifications, the form of the  interactions with the K\"ahler sector is partly determined by the corresponding moduli stabilisation mechanism, which we here don't specify. We show however that the effect of  including the tree-level couplings with the K\"ahler moduli, which appear in a no-scale K\"ahler potential but not in the superpotential, is to induce simple shifts in the spectrum \eqref{eq:spect}. We expect this effect to be relevant also for  fully stabilised compactifications.

This paper is organised as follows: in section \ref{sec:review} we review the form of the ${\cal N}=1$ supergravities arising in the low-energy limit of flux compactifications of type IIB string theory and F-theory. Section \ref{sec:universal} contains our computations and results, and section \ref{sec:concl} our conclusions.

%%%%%%%%%%%%%%%%%%%%%%%%%%%%%%%%%%%%%%%%%%%%%%%%%%%%%%%%%%%
%%%%%%%%%%%%%%%%%%%%%%%%%%%%%%%%%%%%%%%%%%%%%%%%%%%%%%%%%%%
%%%%%%%%%%%%%%%%%%%%%%%%%%%%%%%%%%%%%%%%%%%%%%%%%%%%%%%%%%%

\section{The low-energy theory of flux compactifications 
} \label{sec:review}
 In this section, we briefly review the elements of the four-dimensional effective theories arising from flux compactifications of type IIB string theory and F-theory \cite{Grimm:2004uq, Denef:2008wq}.

\subsection{Type IIB compactifications}
The low-energy spectrum of  compactifications of type IIB supergravity on the orientifold $\tilde M_3$ of the Calabi-Yau three-fold $M_3$
 include the axio-dilaton, $\tau = C_0 + i e^{-\phi}$,  the complex structure moduli, $\phi^i$, where $i=1,\ldots,h^{1,2}_-(\tilde M_3)$, and the K\"ahler moduli, $T^{\alpha}$, where $\alpha=1,\ldots, h^{1,1}_+( \tilde M_3)$.\footnote{ We here ignore the possible existence of  axion multiplets $G^{\alpha}$, with $\alpha = 1, \ldots, h^{1,1}_-(\tilde M_3)$, that are not lifted by the fluxes.}  
We are interested in compactifications in which integrally quantised RR ($F_3$) and NS-NS ($H_3$) fluxes wrap some non-trivial three-cycles of $M_3$, 
\be
\frac{1}{(2\pi)^2 \alpha'} \int_{A^I, B_I} F_3 = \vec{N}_{\rm RR} \in \mathbb{Z}^{2(h^{1,2}+1)}\, ,~~~\frac{1}{(2\pi)^2 \alpha'} \int_{A^I, B_I} H_3 = \vec{N}_{\rm NS-NS} \in \mathbb{Z}^{2(h^{1,2}+1)}\, . \nonumber
\ee
Here  $(A^I, B_I)$  with $I= 0, \ldots , h^{1,2}(M_3)$ denotes an integral and symplectic homology basis of $H_3(M_3, \mathbb{Z})$, satisfying,
\bea
A^I \cap B_J = \delta^I_J \, , ~~~~~~ A^I \cap A^J = B_I \cap B_J = 0 \, .
\eea 
The RR and NS-NS three-form flux is conveniently expressed in the combination $G_3 = F_3 - \tau H_3$, and we denote the corresponding complexified flux vector by,
\be
\vec{N} =  \Sigma \, 
\left(
\begin{array}{c}
 \int_{A^I} G_3 \\
 \int_{B_I} G_3
\end{array}
\right) \, , \qquad \text{with} \qquad \Sigma = 
\left(
\begin{array}{c c}
0 & \mathbb{1} \\
- \mathbb{1} & 0
\end{array}
\right) \, .
\ee

The K\"ahler potential of the  low-energy effective supergravity  of type IIB flux compactifications is given to leading order in $g_s$ and $\alpha'$ by, 
 \be
 K = - \ln \left(i \int_{M_3} \Omega \wedge \bar \Omega \right)
 - \ln \left(-i(\tau - \bar \tau)\right)
 - 2 \ln {\cal V}
 \label{eq:K}
 \, .
 \ee
 Here, ${\cal V}$ denotes the K\"ahler moduli dependent compactification volume and $\Omega$ denotes the complex structure moduli dependent holomorphic three-form of $M_3$. 

The non-trivial fluxes induce a complex structure and axio-dilaton dependent superpotential \cite{Gukov:1999ya},
\be
W =
\int_{M_3}  G_3 \wedge \Omega    
 \, .
\label{eq:GVW}
\ee 
The complex structure dependence is conveniently expressed through the period vector,
\be
\vec{\Pi} =
 \left(
 \begin{array}{c}
 \int_{A^I} \Omega \\
  \int_{B_I} \Omega
  \end{array}
  \right)
\, , 
\label{eq:Period0}
\ee
where $z^I =  \int_{A^I} \Omega$ are homogenous coordinates on the complex structure moduli space. 
The corresponding inhomogeneous coordinates, e.g.~$\phi^i = -i z^i/z^0$ for $z^0\neq0$, defines the corresponding 
complex structure moduli fields   for $i = 1, \ldots, h^{1,2}$.  Upon setting $z^0 =1$, the period vector is given by,  
\be
\Pv = 
\left(
\begin{array}{c}
1 \\
i \phi^i \\
2 F- \phi^j F_j \\
-i F_i
 \end{array}
\right) \, ,
\label{eq:Period}
\ee
where $F(\phi^i)$ denotes the holomorphic ${\cal N}=2$ prepotential of the underlying special geometry of the complex structure moduli space.

The functional form of $F(\phi^i)$ in the neighbourhood of a large complex structure point located at ${\rm Re}(\phi^i) \to \infty$, is given by,
\be
F =  \frac{i}{6} \kappa_{ijk} \phi^i \phi^j \phi^k + \frac{1}{2} \kappa_{ij} \phi^i \phi^j + i \kappa_i \phi^i + \frac{1}{2} \kappa_0 + I \, ,
\label{eq:F}
\ee
where $I$ denotes exponentially suppressed  instanton contributions that will not be important to our  discussion. The classical expansion coefficients are given by intersections of the mirror dual Calabi-Yau three-fold $M_{3}^{\rm d}$  \cite{Hosono:1993qy, Berglund:1993ax, Hosono:1994ax} (see also \cite{Krippendorf}). For historical reasons, the coefficients $\kappa_{ijk}$ are called (the classical) Yukawa couplings \cite{Candelas:1990pi}.

The orientifold action projects out some of the moduli from the spectrum and reduces the amount of supersymmetry to ${\cal N}=1$ in $d=4$.
The complex structure dependent part of the K\"ahler potential may now be written as,
\bea
K_{\rm c.s.} &=& - \ln \left(i \int_{M_3} \Omega \wedge \bar \Omega \right) = 
-\ln \left( i \vec{\Pi}^{\dagger}\, \Sigma\, \vec{\Pi} \right) \nonumber \\
&=&- \ln\left(
\frac{1}{6} \kappa_{ijk}(\phi + \bar \phi)^i(\phi + \bar \phi)^j(\phi + \bar \phi)^k  
 -2 {\rm Im}(\kappa_0)  
\right) \, .
\eea
When,
\be
Y(\phi^i+\bar \phi^i) \equiv \frac{1}{6} \kappa_{ijk}(\phi + \bar \phi)^i(\phi + \bar \phi)^j(\phi + \bar \phi)^k  \gg 2 {\rm Im}(\kappa_0)  \label{eq:Y}
\, ,
\ee 
the K\"ahler potential is well approximated by,
\be
K_{\rm c.s.} \approx
- \ln \left( Y(\phi^i+\bar \phi^i)\right)
\, .
\ee
For any given flux configuration with  $N \equiv (\vec{N})_{h^{1,2}+2} \neq 0$,  it follows from the  LCS expansion \eqref{eq:F}  that the superpotential \eqref{eq:GVW} for sufficiently large  values of the moduli is approximated by,
\be
W = \vec{N} \cdot \vec{\Pi} \approx - i\frac{N}{6} \kappa_{ijk} \phi^i \phi^j \phi^k + {\cal O}(\phi^2) \, .
\ee
Here we have used that the cubic terms arise from a single cycle and  hence appear in the superpotential multiplied by  a single complexified flux component, $N$.

In this paper we only consider  the generic case with $N\neq 0$. For any non-trivial flux configuration, we can choose the inhomogeneous coordinates on moduli space in such a way that the $z^0$ variable corresponds to a cycle with non-vanishing flux, and hence $N \neq 0$. 
%
%there exists at least one LCS point in the moduli space for which this is the case, while there may in general also exist 
Clearly, for a given flux configuration with some vanishing flux numbers, there will in general exist
 LCS points for which the period supporting the cubic terms in the moduli does not appear in the flux superpotential and our results do not apply.

In sum, in this paper we consider the type IIB flux compactifications that are described by the  four-dimensional ${\cal N}=1$ supergravity with, 
\begin{empheq}[box=\fbox]{align}
\vspace*{.5 cm}
K= - \ln (Y) - \ln(-i(\tau - \bar \tau))
 \, , ~~~~ W = - i\frac{N}{6} \kappa_{ijk} \phi^i \phi^j \phi^k \, ,
\label{eq:KW}
\vspace*{.5 cm}
\end{empheq} 
where $Y$ is defined by equation \eqref{eq:Y}.
We will for now  ignore K\"ahler moduli as these do not appear in the flux superpotential, but we will  briefly comment on their inclusion in section \ref{sec:concl}.

%%%%%%%%%%%%%%%%%%%%%%%%%%%%%%%%%%%%%%%%%%%%%%%%%%%%%%%%%%%

\subsection{F-theory compactifications}
Compactifications of F-theory can be viewed as a generalisation of type IIB compactifications to include non-trivial axio-dilaton profiles in the internal space. In this description, $\tau$ is identified with the complex structure modulus of an elliptic curve fibered over a compact three dimensional base manifold (that need not be Calabi-Yau). In supersymmetric compactifications, the resulting four-fold, $X$, is Calabi-Yau  and the four-dimensional ${\cal N}=1$ low-energy supergravity for the complex structure moduli  is described by,
\be
K = - \ln \left( \int_X \Omega_4 \wedge \bar \Omega_4 \right) \, , \qquad \quad W = \int_X G_4 \wedge \Omega_4 \, .
\ee
The complex structure moduli, $\phi^i$ with $i=1, \ldots, h^{3,1}(X)$,  may in analogy with type IIB compactification be defined from the periods of the four-fold. In the large complex structure limit of manifold $X$, one of the periods, say $\Pi_{\rm max}$,  grows as,
\be
\Pi_{\rm max} \sim \kappa_{ijkl} \phi^i \phi^j \phi^k \phi^l \, , 
\ee
while other periods grow at most as $\sim {\cal O}( \phi^3)$. Here $\kappa_{ijkl}$ denote the classical intersection numbers of the divisors of the mirror dual four-fold, as discussed in \cite{Hosono:1994av, Honma:2013hma, Arends:2014qca,Hebecker:2014kva}. 

We will consider flux compactifications of F-theory in which the four-form flux $G_4$ on the corresponding cycle does not vanish. The superpotential and K\"ahler potential are then well-approximated by,
\begin{empheq}[box=\fbox]{align}
\vspace*{.5 cm}
K= - \ln \left(\frac{1}{4!} \kappa_{ijkl} (\phi+\bar \phi)^i (\phi+\bar \phi)^j (\phi+\bar \phi)^k (\phi+\bar \phi)^l \right)  \, , ~~~~ W = N \kappa_{ijkl} \phi^i \phi^j \phi^k \phi^l \, ,
\label{eq:FthKW}
\vspace*{.5 cm}
\end{empheq} 
at large complex structure.  We note that as the Hodge numbers of elliptically fibered Calabi-Yau four-folds tend to be quite large ${\cal O}(h^{3,1}) \sim 10^5$,  the superpotential of equation \eqref{eq:FthKW} is in general a nontrivial function containing  up to $\sim 10^{18}$ terms.\footnote{Our results are equally applicable for the case of sparse Yukawa couplings.} We will nevertheless show that it posses enough structure to be entirely computable with rudimentary techniques.

%%%%%%%%%%%%%%%%%%%%%%%%%%%%%%%%%%%%%%%%%%%%%%%%%%%%%%%%%%%
%%%%%%%%%%%%%%%%%%%%%%%%%%%%%%%%%%%%%%%%%%%%%%%%%%%%%%%%%%%
%%%%%%%%%%%%%%%%%%%%%%%%%%%%%%%%%%%%%%%%%%%%%%%%%%%%%%%%%%%

\section{Universal properties at large complex structure} \label{sec:universal}
In this section we extend the results of \cite{Brodie:2015kza} by showing that  the regions of the moduli space
where the
  low-energy theory is well-described by equation \eqref{eq:KW} or equation \eqref{eq:FthKW}
 exhibit certain universal properties, such as simple expressions for the gradient of the scalar potential  and the spectrum of the Hessian matrix. 

Our method is very similar to that of \cite{Farquet:2012cs} and is quite straight-forward: for a given point $p$ in the moduli space, we make a coordinate transformation $\phi^i \to \varphi^a$ to canonically normalise the K\"ahler metric. In these new coordinates, 
 the Yukawa couplings satisfy certain non-trivial relations which we use to derive the explicit form of the F-terms, the spectrum of the canonically normalised fields, and the inflationary slow-roll parameters.    

\subsection{Type IIB compactifications}

\subsubsection{Canonical normalisation and  Yukawa couplings} \label{sec:cannorm}

We begin by considering a general point $p$ in the  moduli space for which \eqref{eq:KW} provides a good approximation of the low-energy theory.  The K\"ahler metric  is then real  and  in the standard coordinates $(\tau(p), \phi^i(p))$ given by, 
\bea
K_{\tau \bar \tau} &=&- \frac{1}{(\tau - \bar \tau)^2} \, , \\
K_{i \bar \tau} &=& K_{\tau \bar j} = 0 \, , \\
K_{i\bar j}  &=& -\frac{\kappa_{ijk}}{Y}    (\phi^k+\bar \phi^k) 
+ \frac{\kappa_{ilm} \kappa_{jnp}}{4 Y^2} (\phi^l+\bar  \phi^l)  (\phi^m+\bar \phi^m)  (\phi^n+\bar  \phi^n)  (\phi^p+\bar \phi^p) \, .
\label{eq:Kder}
\eea
The metric on the complex structure moduli space can locally at $p$ be brought to the canonically normalised form by a change of variables. Here, we will define ${\cal G} \in {\rm GL}(h^{1,2}_-, \mathbb{R})$ to be the transformation, $\varphi^{a} = {\cal G}^{a}_{~i} \phi^i$, 
that canonically normalises the metric, and  aligns the moduli vacuum expectation value (vev) with the first coordinate axis, such that
$\varphi^a = \varphi_0\, \delta^a_1$.  The K\"ahler potential  and the function $Y$ defined through equation \eqref{eq:Y} are scalars and hence  invariant under this coordinate transformation. The Yukawa couplings  transform as,
\be
\kappa_{abc} = ({\cal G}^{-1})^i_{~ a} ({\cal G}^{-1})^j_{~ b} ({\cal G}^{-1})^k_{~ c}\,  \kappa_{ijk} \, .
\ee
In the new coordinate system, where $K_{a \bar b} = \delta_{a \bar b}$,  equation  \eqref{eq:Kder} reads,
\be
\delta_{a  b} =  -\frac{\kappa_{ab1}}{Y}    x 
+ \frac{\kappa_{a11} \kappa_{b11}}{4 Y^2} x^4 \, , \label{eq:metriceqn}
\ee
where we have denoted $\varphi_0 = x/2+i y/2$, and  dropped the bar on the indices for the real K\"ahler  metric. Note that from the definition \eqref{eq:Y},  $Y = \frac{1}{6} \kappa_{111} x^3$. Equation \eqref{eq:metriceqn} has several non-trivial implications for the  orthonormal-frame Yukawa couplings,
\be
 \kappa_{111} = \frac{2}{\sqrt{3}} \, Y, \qquad \qquad \kappa_{11a'} = 0, \qquad \qquad \kappa_{1a'b'} = - \frac{1}{\sqrt{3}}\, \delta_{a'b'}\,  Y,
\label{eq:kappas}
\ee
in agreement with the results found in \cite{Denef:2004ze,Dimofte:2008jg} for models with a single complex structure modulus. Here $a', b' \in \{2, \ldots, h^{1,2}_-\}$. Crucially important for our following derivations is that  the Yukawa couplings drop out of 
first equation of \eqref{eq:kappas}, which then restricts the value of the real component of $\varphi_0$,
\be
x^2 = 3 \, .
\ee
Since $x>0$, we find that in these coordinates the real part of the modulus vev is uniquely fixed, $\varphi_0 = \sqrt{3}/2 +i y/2$, and $\kappa_{a' b'1} = - \frac{\kappa_{111}}{2} \delta_{a' b'}$. 
 
  The Yukawa couplings $\kappa_{a' b' c'}$ with no index along the `$1$'-direction are not constrained by the canonical normalisation condition, however, we will see these coefficients do not appear in the expressions for the gradient vector or the Hessian matrix. 

The canonical normalisation of the axio-dilaton is achieved by the vielbein,
$e^\tau_0 =i (\tau - \bar \tau)$,  and then we may write  the canonically normalised fields as $\varphi^A$ with $A=0$ corresponding to the normalised axio-dilaton, and $A = a = 1, \ldots , h^{1,2}_-$ for the complex structure moduli.

\subsubsection{Value of the potential and its gradient}
It is well known from mirror symmetry that the K\"ahler potential  for the complex structure moduli close to the LCS point 
is of `no-scale type' satisfying $K_a K^a = 3$.  This follows directly from that the function $Y$ of \eqref{eq:Y} is a homogeneous function of degree 3 in the moduli, and is here explicitly evident as,  in terms of the canonically normalised fields,
\be
K_a \equiv \partial_{\varphi^a} K = - \frac{1}{2} \kappa_{a11} \frac{x^2}{Y} = - \sqrt{3}\, \delta_a^1 \, ,
\ee
so that indeed $K_a K^a = K_a K_{\bar a} \delta^{a \bar a} = 3$. We note that $K_a$ is aligned with the real axis of the complex $\varphi^1$ plane.  

An interesting property of the theory defined by \eqref{eq:KW} is that while $\partial_a W \neq 0$ and $K_a W K^a \overline W = 3|W|^2$, we still find that the squared magnitude of the covariant derivative $F_a = D_aW = (\partial_a +K_a )W$, satisfies $F_a \bar F^a = 3|W|^2$.  To see this, we note that equation \eqref{eq:kappas} implies that,
\be
\partial_a W = \frac{3}{\varphi_0} W \delta_a^1 \, .
\ee
We then have that,
\be
F_a = \frac{\sqrt{3}}{\varphi_0} W \, \delta_a^1\, \left( \sqrt{3} - \varphi_0 \right) = \sqrt{3} W \, \delta_a^1\, \left(\frac{\varphi_0^*}{\varphi_0}  \right) \, , \label{eq:Fa}
\ee
and $F_a \bar F^a = 3|W|^2$. The axio-dilaton F-term is perhaps easiest found by noting that $D_{\tau} N = K_{\tau} N^*$ so that $F_{\tau} \bar F^{\tau} = |W|^2$. 

In sum, for the canonically normalised fields  the F-terms are given by,
\be
F_A = \left(
\begin{matrix}
F_0 \\
F_a
\end{matrix}
\right)
=
\left(
\begin{matrix}
-i \left(\frac{N^*}{N} \right)W \\
\sqrt{3} \left( \frac{\varphi^*_0}{\varphi_0} \right) W \delta_a^1
\end{matrix}
\right) \, .
\label{eq:Fterms}
\ee

Consequently, at a point $p$ in the moduli space for which the K\"ahler potential and superpotential are given by equations \eqref{eq:KW}, the magnitude of the F-terms and the scalar potential are given by,
\be
F_A \bar F^A = 4 |W|^2 \qquad \Longrightarrow \qquad  V = e^K |W|^2 \, . \label{eq:V}
\ee
As we have shown that the second equation in \eqref{eq:V} holds at a generic point at large complex structure, the gradient  can now be found through simple differentiation,
\bea
\partial_A V &=& e^K F_A \overline W \, . \label{eq:Va}
\eea
Near the LCS limit and with $N\neq0$, the superpotential and the F-terms cannot vanish. Thus,  the scalar potential has no critical points in this regime.

\subsubsection{Spectrum}
\label{sec:spectrum}

The Hessian matrix derived from the potential \eqref{eq:V} is given by,
\be
\mathcal{H} = \left(
\begin{array}{cc}
\nabla_{A \bar B} V  & \nabla_{AB} V \\
 \nabla_{\bar A\bar B} V  & \nabla_{\bar A B} V  
\end{array}
\right) =e^{K}  \left(\begin{array}{cc}
  K_{A \bar B}  |W|^2 + F_A \bar F_{\bar B} & \overline W Z_{AB} \\
 W \bar  Z_{\bar A \bar B}&  K_{\bar A  B}  |W|^2 + \bar F_{\bar A}  F_{ B} 
\end{array}
\right)\, ,
\label{eq:Hessian}
\ee
where $Z_{AB} =  D_B F_A = \partial_B F_A +K_B F_A - \Gamma_{AB}^C F_C$. Thus, to find the spectrum in the entire LCS limit, we need only to find  the symmetric complex tensor $Z_{AB}$. The $A=B=0$ component vanishes, as $Z_{\tau \tau} = 0$ due to the linear appearance of $\tau$ in $W$ together with the simple form of the axio-dilaton K\"ahler potential. The mixed components are given by,
\be
Z_{ \tau a} = K_{\tau} \frac{N^*}{N} F_a 
 \qquad \Longrightarrow \qquad Z_{0a} = e_0^{\tau} Z_{\tau a} = 
-i   \sqrt{3} |W| \, e^{i \theta} \, \delta_a^1
\, ,
\ee
where we have defined the phase,\footnote{Note that these phases are related by the identity, $W/\overline W = - (N/N^*) (\varphi_0/\varphi_0^*)^3$.} 
\be
\theta = 
{\rm arg} \left(W
  \left( \frac{N^*}{N} \right) \left(\frac{\varphi_0^*}{\varphi_0} \right) 
\right)  \, . 
\ee
The remaining components can be determined from the identity,
\be
Z_{ij} = -(\tau -\bar \tau) e^{K_{\rm c.s.}} \kappa_{ij}^{~~\bar k}\, \bar Z_{\bar \tau \bar k} \, , 
\ee
 as reviewed in \cite{Brodie:2015kza}. For  canonically normalised fields this expression reads, 
 \be
 Z_{ab} = i\, e^{K_{\rm c.s.}}  \kappa_{ab}^{~~\bar c}\, \bar Z_{\bar 0 \bar c} = - \frac{\sqrt{3}}{Y} \kappa_{1ab} |W| e^{-i \theta} 
 \, ,
 \ee
and the   tensor $Z_{AB}$ can be expressed in matrix representation as,
\be
Z_{AB} = 
\left(
\begin{matrix}
Z_{00} & Z_{01} & Z_{0b'}  \\
Z_{10} & Z_{11} & Z_{1 b'} \\
Z_{a' 0} & Z_{a' 1} & Z_{a' b'} 
\end{matrix}
\right)
=
|W|
\left(
\begin{matrix}
0 & - \sqrt{3} i e^{i \theta} & 0  \\
- \sqrt{3} i  e^{i \theta} & -2 e^{-i \theta} & 0 \\
0 & 0 &  \delta_{a' b'} e^{-i \theta} 
\end{matrix}
\right) \, .
\label{eq:Zmat}
\ee

We are now ready to deduce the spectrum of the axio-dilaton and complex structure moduli sector at LCS. First, we note that the $(a', b')$ directions have no cross-terms with the axio-dilaton or the  `1'-direction, and hence decouple. 
The Hessian matrix in this sector is given by $h^{1,2}_--1$ copies of the same $2\times2$ matrix, 
\be
 {\cal H}_{\perp}
=
e^K |W|^2
\left(
\begin{matrix}
1 & e^{-i( \theta+ \theta_W)} \\
e^{i (\theta +\theta_W)} & 1 
\end{matrix}
\right) \, ,  
\ee
where $\theta_W = {\rm arg} \left(W\right)$, and the corresponding eigenvalues are given by, 
\bea
m^2_{a'\pm} = (1\pm1)\, m_{3/2}^2 \, , \label{eq:mapm}
\eea
for $a' = 2, \ldots, h^{1,2}_-$.

The remaining eigenvalues and eigenvectors are perhaps easiest found by first noting that the F-terms \eqref{eq:Fterms} satisfy the equation,
\be
Z_{AB} \bar F^B = 3 \overline W F_A \, ,
\ee
with $Z_{AB}$ given by equation \eqref{eq:Zmat}.
  The Hessian matrix then has the following eigenvectors and  associated eigenvalues, 
  \be
  v_{F\pm} = \left(
\begin{matrix}  e^{-i \theta_W }F^{\bar B} \\ \pm e^{i \theta_W } \bar F^B
\end{matrix}
\right) \,  , \qquad m^2_{F\pm} = (5 \pm 3)\, m_{3/2}^2 \, . \label{eq:mfpm}
  \ee
 
 The final eigenvectors correspond to  linear combinations of the `0' and `1' directions that are perpendicular to the vectors $v_{F\pm}$. These can be written in   the form, 
\be
 v_{h\pm} = \left(
\begin{matrix}
e^{-i \theta_W}h^{\bar B} \\
 \pm e^{i \theta_W} \bar h^B
\end{matrix}\right) \, ,
 \ee
where $\bar h^B F_B = 0$ and, 
\be
  Z_{AB} \bar h^B =  \overline  W h_A \, .
\ee
Explicitly we find,
\be
h_A =
\left(
\begin{matrix}
- \sqrt{3} i \left( \frac{N^*}{N} \right)^{1/2} \left( \frac{\varphi_0}{\varphi_0^*}\right)^{3/2}\\
- \left( \frac{N}{N^*} \right)^{1/2} \left( \frac{\varphi_0}{\varphi_0^*} \right)^{1/2}
\end{matrix}
\right)
=
\exp\left(\frac{i}{2}( \theta_W - \theta) \right)\, \left(
\begin{matrix}
 \sqrt{3} i\,  e^{2 i \theta} \\
-\,  \delta_a^1
\end{matrix}
\right)
\, ,
\ee
with the eigenvalues of the Hessian matrix,
\bea
m^2_{h\pm} &=& (1\pm1)\, m_{3/2}^2 \, . \label{eq:m1pm}
\eea

Thus, by equations \eqref{eq:mapm}, \eqref{eq:mfpm} and \eqref{eq:m1pm}, we have shown that, for the entire region of the LCS expansion in which the theory is well-approximated by \eqref{eq:KW},  the spectrum is given by equation 
\eqref{eq:spect}. This is the main result of this paper.

We close this section by commenting on the spectrum of the $2 (h_-^{1,2}+1)$ dimensional hermitian matrix ${\cal M}$, which is constructed as, 
\be
{\cal M} = 
e^{\frac{K}{2}}\left(
\begin{matrix}
0 & Z_{AB} e^{- i\theta_W } \\
\bar Z_{\bar A \bar B}  e^{i\theta_W}& 0
\end{matrix}
\right)
\, .
\ee
The matrix ${\cal M}$ governs the non-supersymmetric critical point equation \cite{Denef:2004cf}, and its eigenvalues  determine the  spectrum of fermion masses at vacua of the scalar potential and  the supersymmetric contribution to the scalar masses. From equation \eqref{eq:Zmat}, we find that its spectrum is given by,
\be
\lambda_{i\pm} =
\left\{
 \begin{array}{l c l }
 \pm \, m_{3/2} &~~& i=1, \ldots,h^{1,2}_-, \\
\pm \, 3\, m_{3/2} &~~& i=h^{1,2}_-+1  \, , 
\end{array}
\right.
\ee
which again generalises the findings of  \cite{Brodie:2015kza}. 

%%%%%%%%%%%%%%%%%%%%%%%%%%%%%%%%%%%%%%%%%%%%%%%%%%%%%%%%%%%

\subsection{F-theory compactifications}
The computation of the values of the potential, its gradient and the spectrum of F-theory compactifications that are well-described by  \eqref{eq:FthKW} proceed  very similarly to the type IIB case. This is not surprising, as equation \eqref{eq:FthKW} can be viewed as a special case of \eqref{eq:KW} with negligible RR-flux  on the cycle supporting  a period with  cubic dependence on the complex structure moduli, upon identification of the axio-dilaton as an additional complex structure modulus. However, the computation of the properties of the low-energy theory in F-theory is perhaps even more transparent than the type IIB case, as we  now show.

\subsubsection{Canonical normalisation and Yukawa couplings}
The K\"ahler metric derived from equation \eqref{eq:FthKW} is given by,
\bea
&K_{ij}  =& -12 \frac{\kappa_{ijkl} (\phi+\phi^*)^k (\phi+\phi^*)^l }{\kappa_{mnop} (\phi+\phi^*)^m (\phi+\phi^*)^n(\phi+\phi^*)^o (\phi+\phi^*)^p} 
\nonumber \\
&+ 16&\frac{\kappa_{i klm}  \kappa_{j  nop} (\phi+\phi^*)^k (\phi+\phi^*)^l(\phi+\phi^*)^m (\phi+\phi^*)^n(\phi+\phi^*)^o (\phi+\phi^*)^p}{(\kappa_{qrst} (\phi+\phi^*)^q (\phi+\phi^*)^r(\phi+\phi^*)^r (\phi+\phi^*)^t)^2} \, .
\eea
Again, the K\"ahler metric is real and can locally be brought into canonical form by a ${\cal G} \in {\rm GL}( h^{3,1}(X), \mathbb{R})$ transformation acting as $\varphi^{a} = {\cal G}^{a}_{~i} \phi^i$,
 where $i$ and $a$  now run from $1$ to $h^{3,1}(X)$. As in section \ref{sec:cannorm}, we take the transformation ${\cal G}$ to rotate the field vevs to align with the first coordinate axis, $\varphi^a = \varphi_0\, \delta^a_1$.

Expressed in  the canonically normalised coordinates, the metric gives rise to the following non-trivial relations,
\be
x^2 = 4 \, , ~~~~~~~~ \kappa_{a' 111} =0 \, , ~~~~~~~~ \kappa_{a' b' 1 1} = - \frac{\kappa_{1111}}{3}\,  \delta_{a' b'} \, .
\label{eq:Fthxeqn}
\ee 
for $a', b' \in \{2, h^{3,1}(X) \}$. The first equation of \eqref{eq:Fthxeqn} implies that  $0< x = \varphi_0 + \varphi_0^* = 2$.

\subsubsection{Value of the potential and its gradient}
The K\"ahler potential of F-theory compactifications at LCS is given by the logarithm of a function that is quartic in the moduli vevs -- not cubic as in the type IIB case -- and hence these compactifications  are not no-scale at large complex structure. In canonically normalised coordinates, we see this explicitly  as $K_a = -2 \delta_a^1$ and $K_a K^a = 4$.  Using \eqref{eq:FthKW} and \eqref{eq:Fthxeqn}, we find the F-terms,
\be
F_a = W \left( \frac{4}{\varphi_0} -2 \right)\, \delta_a^1 =2W \left(\frac{\varphi_0^*}{\varphi_0} \right)\, \delta_a^1 \, ,
\ee
so that $F_a \bar F^a = 4|W|^2$. The values of the potential and its gradient are given by,
\be
V = e^K |W|^2\, , ~~~~~ \partial_a V = e^K F_a \overline W \, , \label{eq:FthV}
\ee
just as equations  \eqref{eq:V} and \eqref{eq:Va} for the type IIB compactifications. 

\subsubsection{Spectrum}
The Hessian matrix of the  system described by \eqref{eq:FthKW} is given by \eqref{eq:Hessian}. Thus, we need only to compute $Z_{ab} = D_a F_b = \partial_a F_b + K_a F_b - \Gamma_{ab}^c F_c$. The relevant field space Christoffel symbols are very simple,
\be
\Gamma^1_{ab} = - K_{ab} \, .
\ee
Again using that $2/\varphi_0 -1 = \varphi^*_0/\varphi_0$, we express 
the components of $Z_{ab}$ as,
\bea
Z_{11} &=& 3 W \left( \frac{\varphi^*_0}{\varphi_0}\right)^2 \, , \\
Z_{1b'} &=& 0 \, , \\
Z_{a'b'} &=& - \delta_{a'b'} W  \left( \frac{ \varphi^*_0}{\varphi_0}\right)^2 \, .
\eea
The Hessian matrix is now simply given by $h^{3,1}(X)$ decoupled $2\times 2$ matrices. The `1'-direction is special and the corresponding elements of the Hessian matrix are given by,
\be
{\cal H}_F =
e^K
\left(
\begin{matrix}
\nabla_{1 \bar 1} V & \nabla_{1 1} V \\
\nabla_{\bar 1 \bar 1} V &  \nabla_{\bar 1 1} V  
\end{matrix}
\right)
=e^K |W|^2
\left(
\begin{matrix}
5 & 3 \left( \frac{ \varphi^*_0}{\varphi_0}\right)^2 \\
3\left( \frac{ \varphi_0}{\varphi^*_0}\right)^2  & 5
\end{matrix}
\right) \, ,
\ee
with the eigenvalues,
\bea
m^2_{F_\pm} = (5\pm 3)\, m_{3/2}^2 \, .
\eea
The remaining eigenvalues of the Hessian are given by $h^{3,1}(X)-1$ copies of the  spectrum of the $2\times 2$ matrix,
\be
{\cal H}_{\perp} =
e^K |W|^2
\left(
\begin{matrix}
1 & - \left( \frac{ \varphi^*_0}{\varphi_0}\right)^2 \\
-\left( \frac{ \varphi_0}{\varphi^*_0}\right)^2  & 1
\end{matrix}
\right) \, ,
\ee
giving,
\be
m^2_{a'\pm} = (1\pm1)\, m_{3/2}^2 \, .
\ee
Thus, in sum, the spectrum of  complex structure moduli of  F-theory compactifications  at LCS, as described by the low-energy theory \eqref{eq:FthKW}, is exactly given by equation \eqref{eq:spect} upon substituting $h^{1,2}_-(\tilde M_3) \to h^{1,3}(X)$.

%%%%%%%%%%%%%%%%%%%%%%%%%%%%%%%%%%%%%%%%%%%%%%%%%%%%%%%%%%%
%%%%%%%%%%%%%%%%%%%%%%%%%%%%%%%%%%%%%%%%%%%%%%%%%%%%%%%%%%%
%%%%%%%%%%%%%%%%%%%%%%%%%%%%%%%%%%%%%%%%%%%%%%%%%%%%%%%%%%%

\section{Conclusions} \label{sec:concl}
We have shown, in generalisation of the results of \cite{Brodie:2015kza}, that flux compactifications close to large complex structure points exhibit universal properties that are independent of the underlying details of the compactification. In the regions of the moduli space where the flux superpotentials of type IIB and F-theory compactifications are dominated by cubic or quartic terms in the complex structure moduli, respectively, the energy density is positive definite, there are no vacua, and the spectrum of the Hessian consists merely of  three distinct eigenvalues. 

These results should be contrasted with the expectations from the continuous flux approximation \eqref{eq:contFlux}, which for large values of the D3-tadpole predicts an independence of the distribution of the superpotential and the F-terms at each point in the moduli space (c.f.~e.g.~section 5.4 of \cite{Brodie:2015kza}).  Here we find that the magnitude of the F-terms is completely determined by the  value of $|W|$ (c.f.~equation \eqref{eq:V}), and consequently,  the continuous flux approximation does not apply. `Eigenvalue repulsion' and hence a spectrum with non-degenerate eigenvalues is a key feature of random matrix models. We here find that the spectrum typically consists of highly degenerate eigenvalues, and hence, standard random matrix theory techniques are not applicable to flux compactifications in this limit.

The absence of  supersymmetric vacua near the LCS limit of general Calabi-Yau compactifications with generic fluxes is particularly relevant for studying the finiteness of the number of supersymmetric vacua \cite{Ashok:2003gk,Denef:2004ze,Acharya:2006zw,Eguchi:2005eh,Torroba:2006kt,Bousso:2000xa,Magda2}. Indeed,  it was argued in \cite{Ashok:2003gk} that infinite sequences of supersymmetric vacua can only occur if they accumulate in a neighbourhood of `D-limits',   which are singular points of the complex structure moduli space
in which  $\{\vec{\Pi}, \vec{\Pi}^*, D_a \vec{\Pi}, \bar D_{\bar a} \vec{\Pi}^*\}$ degenerates and ceases to be a good symplectic basis. The LCS  point is the canonical example of a `D-limit'. 

The actual existence of such an infinite sequences of vacua was left as an open question in \cite{Ashok:2003gk}, and in  \cite{Eguchi:2005eh,Torroba:2006kt} the statistical methods developed in \cite{Ashok:2003gk} (based on the continuous flux approximation) were used to argue that no such sequence appears close to LCS points. Moreover, for one parameter Calabi-Yau compactifications  it was shown in \cite{Magda2} that the neighbourhood of an LCS point contains no supersymmetric vacua, except  for one sitting right at the LCS point \cite{Danielsson:2006xw}. 

We have here shown that the continuous flux approximation, and hence the statistical arguments of \cite{Ashok:2003gk, Eguchi:2005eh, Torroba:2006kt}, are not applicable for generic fluxes close to a LCS point of a general flux compactification. Nevertheless, for these generic fluxes there are no vacua (and in particularly no infinite sequence of vacua) at large complex structure, in agreement with earlier statistical studies. Our results are consistent with the type IIB vacua found in \cite{Danielsson:2006xw,Magda2} in which  the cubic term in the superpotential  vanishes.   It would be interesting to extend our results to  the non-generic case in which the superpotential is dominated by quadratic terms in the moduli.

In this paper we have considered the spectrum of moduli explicitly lifted by the flux-induced superpotential, i.e.~the complex structure moduli and the axio-dilaton. It is straightforward to extend this analysis to include K\"ahler moduli with the no-scale  K\"ahler potential $K_{(K)} =-2\ln {\cal V}$. In type IIB compactifications, ${\cal V}$ is a homogeneous function of degree $3/2$ in the four-cycle volumes of the compactification three-fold. The corresponding K\"ahler moduli F-term is given by $F_{\alpha} = K_{\alpha} W$, and the scalar potential and gradient now becomes,
\be
V = 4 e^K |W|^2 \, , ~~~~~~\partial_A V = 4 e^K F_A \overline W \, ,
\ee
where $A$ now runs over all moduli, including the $h^{1,1}(\tilde M)$ K\"ahler moduli. The spectrum of the Hessian matrix can be found following a similar method to that of section \ref{sec:spectrum}, and is given by,
\be
m^2_{i} =
\left\{
 \begin{array}{l c l }
 0 &~~& i=1, \ldots,h^{1,2} +h^{1,1}, \\
  8\, m_{3/2}^2 &~~& i=h^{1,2}+h^{1,1} + 1, \ldots, 2(h^{1,2}+h^{1,1})+1, \\
  56\, m_{3/2}^2 && i= 2(h^{1,2} +h^{1,1}+1) \, . \, 
\end{array}
\right.
    \label{eq:fullSpect} 
\ee
To stabilise the K\"ahler moduli additional contributions, such as e.g.~non-perturbative superpotential corrections,  need to be included in the theory, and equation \eqref{eq:fullSpect} would only give the approximate spectrum in regions of moduli space in which such corrections are small. 

Our results can be used to compute the inflationary slow-roll parameters in the regions of moduli space in which \eqref{eq:KW} or \eqref{eq:FthKW} apply. Consistently with \cite{Brodie:2015kza}, we find that (ignoring K\"ahler moduli),
\be
\epsilon = \frac{1}{2} \left( \frac{2\partial_A V K^{A \bar B}\partial_{\bar B} V}{V^2}\right) = 4 \, , ~~~~~~~~~ \eta_{\parallel} = \frac{{\bf e} \cdot {\cal H}\cdot  {\bf e}}{V} = 8 \, ,
\ee
where ${\bf e}$ denotes the unit vector in the $(\partial_A V, \partial_{\bar A} V)^{\rm T}$ direction. Including K\"ahler moduli with the no-scale K\"ahler potential does not qualitatively change this result, as the slow-roll parameters then are given by,
\be
\epsilon  = \frac{7}{4} \, , ~~~~~~~~~ \eta_{\parallel}  = 14 \, .
\ee
We believe that these results are highly relevant for proposed attempts to construct models of axion-monodromy inflation around  LCS points in the moduli space \cite{Garcia-Etxebarria:2014wla}.

There are a number of interesting future directions of this work: while we have shown that general and typically very complicated flux compactifications exhibit universal properties in some region of the moduli space, it would be most interesting to determine if there exist classes of \emph{vacua} that exhibit some universal properties as well.  Moreover, it would be interesting to extend these results to other `special points' in the moduli space, such as conifold points, Landau-Ginzburg points, and geometric engineering limits. We hope to return to these questions in future work.

\section*{Acknowledgments}
We would like to thank Jose Juan Blanco-Pillado, Andreas Braun, Roberto Valandro and Timo Weigand  for stimulating discussions. KS acknowledges financial support from the Spanish Consolider-Ingenio 2010 program CPAN CDS2007-00042, the Consolider EPI CSD2010-00064, the Basque Government (IT-559-10), the Spanish Ministry of Science grant (FPA 2012-34456) and by the ERC Advanced Grant 339169 ``Selfcompletion''.

\bibliographystyle{JHEP}
\bibliography{refs}

\end{document}